\documentclass[usenatbib,useAMS]{mnras}
\usepackage[T1]{fontenc}
\usepackage{ae,aecompl}
\usepackage{graphicx}
\usepackage{amsmath}
\usepackage{amssymb}

\title[{\it RXTE} observations of 2S~1417-624]
{Spectral and timing studies of 2S~1417-624 during a giant outburst}

 \author[Gupta et al.] {Shivangi Gupta$^{1}$\thanks{E-mail: shivangi@prl.res.in}, 
Sachindra Naik$^{1}$\thanks{snaik@prl.res.in}, Gaurava K. Jaisawal$^{2}$\thanks{gaurava@space.dtu.dk},
Prahlad R. Epili$^{1}$\thanks{prahlad@prl.res.in}\\
$^1$ Astronomy and Astrophysics Division, Physical Research Laboratory, Navrangapura, Ahmedabad - 380009, Gujarat, India\\
$^2$ National Space Institute, Technical University of Denmark, Elektrovej 327-328, DK-2800 Lyngby, Denmark}
\begin{document}

\date{}

\maketitle

\begin{abstract}

We present the results obtained from timing and spectral studies of the accretion
powered X-ray pulsar 2S~1417-624 during a giant outburst in 2009 by using {\it Rossi 
X-ray Timing Explorer (RXTE)} observations. X-ray pulsations were detected in the 
light curves obtained from all epochs of observations. The pulsar was found to be 
spinning-up during the outburst. The pulse profiles were observed to be strongly 
dependent on photon energy and luminosity. A double peaked profile at lower luminosity 
evolved into a triple peaked profile at the peak of the outburst which is further 
reverted  back to a double peaked structure during the decay of the outburst. An 
anti-correlation was also observed between the pulse fraction and the source flux. 
The 3-70 keV energy spectrum of pulsar was well described with a power law modified 
with high energy cutoff model along with an iron fluorescence line at 6.4 keV. Based 
on the evolution of pulse profile, pulse fraction and spectral parameters across 
observed luminosity, we interpret our results in terms of changes in the pulsar 
beam configuration from sub-critical to super-critical regimes.


\end{abstract}


\section{Introduction}

Accretion powered X-ray pulsars were discovered in early seventies with {\it Uhuru} 
satellite soon after the long awaited discovery of neutron star in radio band 
(\citealt{Hewish1968,Giacconi1971}). These sources are accreting neutron stars 
that appeared to be one of the brightest X-ray objects in the sky due to their 
transient activity. Most of these sources belong to the class of high mass X-ray 
binaries (HMXBs), where a magnetized neutron star with magnetic field in the order
of B $\sim$ 10$^{12}$~G accretes matter from a supergiant or non-giant optical companion. 
Among HMXBs, Be/X-ray binaries represent about two-thirds of their population. The 
optical companion in these systems is a non-supergiant OB star that shows emission 
lines in optical/infrared spectra at some point of the evolution \citep{Reig2011}. 
The presence of an equatorial circumstellar disk around the Be star is known to be 
the cause of the emission lines in the optical/infrared spectrum and observed infrared 
excess compared to the classical B stars (\citealt{Porter2003, Reig2011}). The orbiting 
neutron star in the Be/X-ray binary system captures significant amount of matter from 
the circumstellar disk at the periastron passage leading to X-ray outbursts (Type~I) 
lasting for a few days to a few weeks. The peak luminosity during these events reaches 
up to one or two order of magnitude higher ($\sim$10$^{35-37}$~erg s$^{-1}$) than the 
quiescent luminosity \citep{Stella1986}. Another kind of X-ray enhancement i.e. Type~II 
outbursts are also observed from neutron stars in Be/X-ray binaries. These outbursts are 
rare and independent of orbital phases of the binary system. These outbursts usually cover 
a significant fraction of the orbit(s) and last for several weeks to months with a peak 
luminosity of $\sim$10$^{38}$~erg s$^{-1}$ (\citealt{Negueruela1998, Paul2011}).  

The transient Be/X-ray binary pulsar 2S~1417-624 was discovered in 1978 with 
{\it SAS-3} \citep{Apparao1980}. During early observations, X-ray pulsations at 
$\sim$17.5~s were detected in the source light curves \citep{Kelley1981}. A B-type 
star located at a distance of 1.4-11.1 kpc and within the search box provided by 
{\it Einstein} observatory was identified as the optical companion of the pulsar 
\citep{Grindlay1984}. Using {\it BATSE} monitoring data of six consecutive outbursts 
between 1994 August and 1995 July, the binary orbital parameters of the system such 
as orbital period $P_{orb}$=42.12 d, eccentricity $e$=0.446, semi major axis $a_x~sin~i$=188~lt-s, 
$\omega$=300$^\circ$.3 and time of periastron passage $T$= JD 2449714.12 were derived 
by \citet{Finger1996} (see also \citealt{Raichur2010}). These authors also reported a 
spin-up of the pulsar at a rate of (3-6) $\times$ 10$^{-11}$ Hz~s$^{-1}$ and found a 
correlation between spin-up rate and pulsed flux. 

After almost four years of quiescence, 2S~1417-624 went into a strong X-ray outburst   
in 1999 November which was subsequently followed by a series of four mini-outbursts 
between 1999 December and 2000 August. Using {\it RXTE} observations of the pulsar 
during these outbursts, intensity dependent pulse profiles and pulsed fraction were 
reported \citep{Inam2004}. The pulsar continuum spectrum in 3-20 keV range was 
described with a power law model modified with cutoff at higher energies. Observed 
variations in spectral and timing parameters were explained in terms of disc accretion 
except at low flux durations where a temporary accretion geometry change was speculated 
\citep{Inam2004}. A complex iron line in 6.4-6.8 keV range was also detected during 
1999-2000 {\it RXTE} observations of the pulsar. Detection of complex iron line indicated 
that the emission was originated from cold iron atoms along with H or He-like ionized gas 
around the neutron star \citep{Inam2004}. A {\it Chandra} observation of the pulsar in 
quiescent phase showed a simple continuum spectrum that was described by a power law or 
a blackbody component at a temperature of 1.5 keV \citep{Tsygankov2017}. Unusual high 
blackbody temperature suggests that the neutron star was accreting from a stable cold 
disk in the faint state.  Recently, the pulsar was detected in a giant outburst 
in 2018 March at a flux of $\sim$350 mCrab in 15-50 keV range  (\citealt{Nakajima2018, 
Krimm2018}). Prior to the 2018 outburst, a giant outburst was also detected in 
2009 November with a peak intensity of $\sim$300~mCrab with {\it Swift}/BAT 
(Fig.~\ref{fig1}; \citealt{Krimm2009}). The pulsar was monitored at multiple 
epochs with the {\it RXTE} during this giant outburst. Using these observations, 
we investigate the evolution of pulse profiles, accretion geometry and spectral 
shape of the pulsar during the 2009 outburst. The present paper describes the 
observation details and analysis methods in Section~2. A description on timing 
and spectral studies are presented in Section~3, followed by result and discussion 
in next section.

\section{Observations and Analysis}

The {\it Rossi X-ray Timing Explorer} ({\it RXTE}) was  launched 
on 1995 December 30 in a low earth orbit. It monitored the X-ray sky 
extensively for about 16 years before being decommissioned in 2012 January. 
The {\it RXTE} consisted of three sets of instruments: All Sky Monitor 
({\it ASM}; \citealt{Levine1996}), Proportional Counter Array ({\it PCA}; 
\citealt{Jahoda1996}) and High Energy X-ray Timing Experiment (HEXTE; 
\citealt{Rothschild1998}), that provided a broad-band coverage in 3-250 keV 
range. The {\it ASM} was sensitive in 1.5-12 keV range. The {\it PCA} unit 
consisted of five identical Xenon filled proportional counters operated in 
2-60 keV energy range. The total effective area of PCA was $\sim$6500~cm$^2$ 
at 6 keV. The PCA has an energy resolution of $<$18\% at 6 keV along with a 
timing resolution of 1$\mu$s. The third instrument, {\it HEXTE} was operated 
in the hard X-ray range of 15 to 250 keV. The {\it HEXTE} comprised of two 
set of clusters called A and B, rocking orthogonal to each other for the 
simultaneous measurement of source and background. Each cluster was made up 
of four NaI(Tl)/CsI(Na) phoswich scintillation counters with a total collecting 
area of $\sim$1600~cm$^{2}$. 

\begin{figure}
\centering
\includegraphics[height=3.2in, width=2.3in, angle=-90]{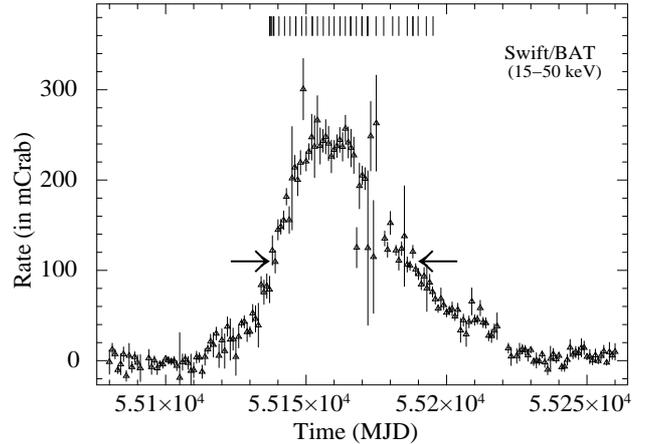}
\caption{{\it Swift}/BAT monitoring light curve of 2S~1417-624 in 15-50 keV range
during a giant outburst in 2009. Arrow marks indicate the beginning and end of 
the {\it RXTE} pointed observations. A total of 39 {\it RXTE} pointed observations 
were made during this period and indicated by the vertical lines at the top of the 
figure.}
\label{fig1}
\end{figure}

We used a total of 39 pointed {\it RXTE} observations of the pulsar 2S~1417-624
carried out between 2009 November 2 to 2009 December 30. This covers
an effective exposure of $\sim$133~ks during the 2009 giant X-ray outburst. 
The dates of these observations are marked by vertical lines across the 15-50 keV 
light curve of the pulsar (Fig.~\ref{fig1}) obtained from the {\it Swift}/BAT 
monitoring data (Krimm et al. 2013). Two additional horizontal arrows 
indicate the beginning and end of {\it RXTE} pointed observations of the pulsar. 
The log of the observations is presented in Table~\ref{log}.

We used Standard-1 and Standard-2 binned mode data from PCA detectors 
to investigate the pulse profile and spectral evolution of the pulsar. 
For this, {\tt HEASoft} package of ver 6.16 was utilized along with the  
calibration data base of {\it RXTE}. We first created a good time interval 
file by applying filter selections on all available PCUs on individual 
observation. The filter selection was done by considering the electron
contamination of $<$0.2, offset angle of $<$0.02 and elevation angle of 
$>$10 and other standard parameters. Using {\it saextrct} task, source light 
curve was extracted in 2-60 keV range at a time resolution of 0.125~s 
from Standard-1 data. Corresponding background light curve was also 
generated from Standard-2 data by using background model provided by the 
instrument team. Barycentric correction was applied on the background 
subtracted light curves to incorporate the motion of satellite and Earth to the 
barycenter of the solar system by using {\it faxbary} task of {\tt FTOOLS}.
For energy resolved light curves, we considered GoodXenon data by applying 
{\it make\_se} task. The Standard-2 (PCA) and Cluster-B (HEXTE) mode data 
were used for carrying out spectral analysis. Phase-resolved spectroscopy was 
also performed by using GoodXenon data that have 256 energy channels. The 
response matrices were created by following the standard procedures. 

\begin{table}
\caption{Log of {\it RXTE} observations of the pulsar 2S~1417-624 during 2009 outburst.}
\begin{tabular}{cccc}
\hline\\
 Proposal    &Number of		 &Time range       &Exposure Time \\

 ID          &Obs. (IDs)   	 &(MJD)         	 &(ksec)\\
\hline
\\
94032            &29  		 &55137.08--55171.96    &98.52 \\
94444            &10    	&55177.64--55195.20    &34.32 \\
\hline\\
Total		&39		&55137--55195		&132.84 \\
\\
\hline
\end{tabular}
\label{log}
\end{table}

\begin{figure*}
\centering
\includegraphics[height=6.5in, width=4.2in, angle=-90]{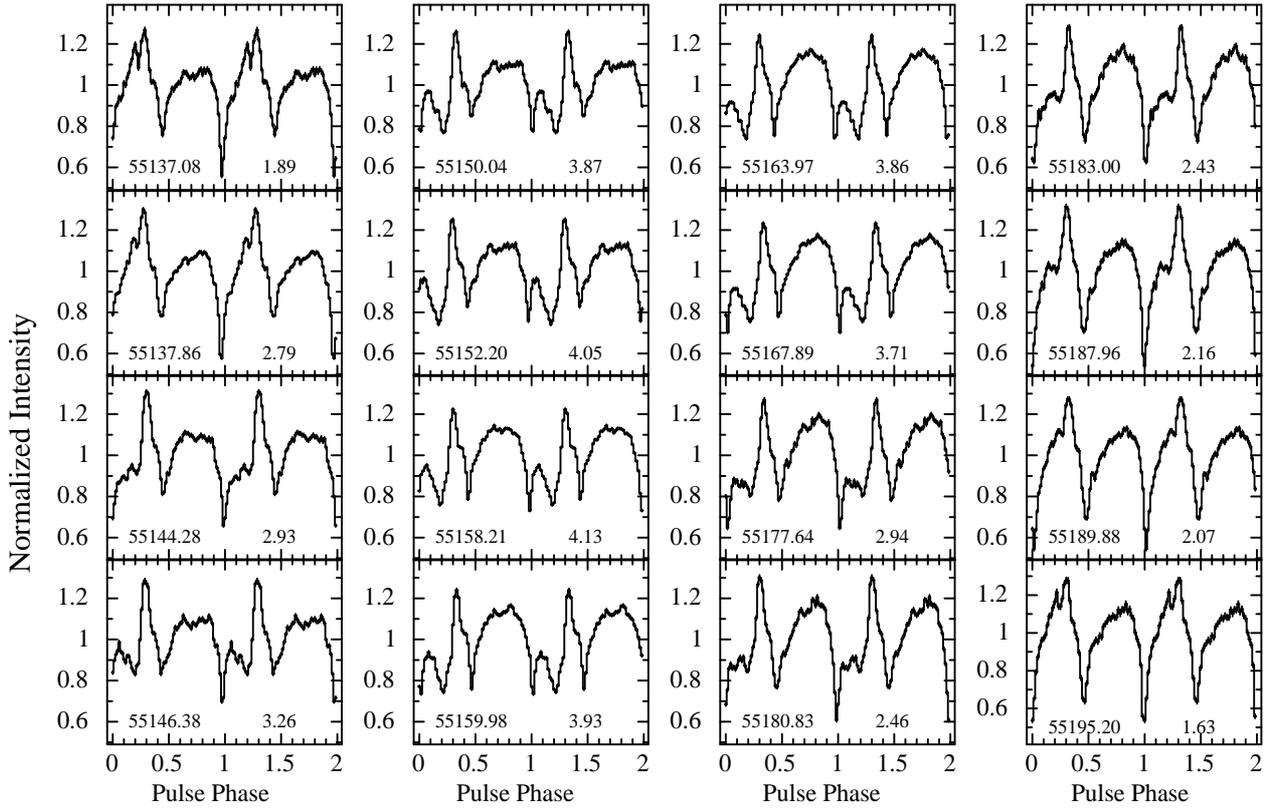}
\caption{Pulse profiles of the pulsar at different phases of the 2009 giant X-ray outburst 
(from rise to decline). The profiles were generated by folding the 2-60 keV light curves 
from PCA data at the respective spin periods. The numbers quoted on the left and right side of 
each panel indicate the beginning of the corresponding {\it RXTE} pointed observation (in MJD) 
and the flux (in 10$^{-9}$~erg s$^{-1}$ cm$^{-2}$ units), respectively. Two pulses are shown in 
each panel for clarity. The error-bars represent 1$\sigma$ uncertainties.} 
\label{fig2}
\end{figure*}

\section{Results}

\subsection {Timing analysis}

Source and background light curves were extracted from all the {\it RXTE/PCA} 
pointed observations at a time resolution of 0.125~s by following procedures 
as described in Section~2. Barycentric correction was also applied on the 
background subtracted light curves. Thereafter, the pulsation period of the 
pulsar was estimated by using a $\chi^2$-maximization technique ({\it efsearch} 
task of {\tt FTOOLS};  \citealt{Leahy1987}). Following this procedure, the barycentric
corrected spin frequency of the pulsar was obtained from the light curves of all
pointed observations and found to be in the range of 57.1-57.25~mHz, showing a 
clear spin-up trend during the 2009 giant X-ray outburst.   

Pulse profiles were generated by folding the light curves at respective 
spin period of the pulsar. This was done to understand the evolution of 
the pulse profile with luminosity and as the outburst progressed. While
generating the pulse profiles for all the pointed observations, we chose 
the folding epoch close to the start time of the respective observation 
and then adjusted in such a way that the minima of the pulse profile align 
at phase zero. Some of the representative pulse profiles in 2-60 keV 
range are shown in Fig.~\ref{fig2}. These profiles are arranged as the
outburst progressed. The numbers quoted in each panel of the figure represent
the beginning of the {\it RXTE} observation (in MJD; left) and the estimated 
flux in 3-30 keV range from spectral fitting (see Section~3.2 and Table~\ref{table2}; 
right). A careful investigation of the shape of the pulse profiles shown in 
Fig.~\ref{fig2} indicated a systematic evolution with the source flux. At 
lower intensity, the pulse profiles appeared as double peaked. Among these, 
the peak appearing in 0-0.5 phase range evolved and gradually splitted into 
two components as the outburst progressed towards the peak (MJD 55146-55167). 
This effectively produces a triple peaked profile at a flux level of 
$\ge$3.2$\times$10$^{-9}$~erg s$^{-1}$~cm$^{-2}$. Assuming the
source distance as 11 kpc (\citealt{Grindlay1984, Inam2004}), corresponding 
3-30 keV luminosity of the pulsar was estimated to be 4.6$\times$10$^{37}$ 
erg s$^{-1}$. The pulsar then restored back to its double peaked pulse profile 
shape as the outburst faded. We also noticed that the shape of profile was 
consistent at comparable luminosities irrespective of rising and declining 
phases of the outburst (see Fig.~\ref{fig2}). Though a double peaked structure 
of the pulse profile has been already seen in 2S~1417-624 (\citealt{Finger1996, 
Inam2004}), the evolution of double-peaked profile to triple-peaked profile 
makes the present study interesting to explore it further through spectral studies.  

\begin{figure}
\centering
\includegraphics[height=3.2in, width=3.9in, angle=-90]{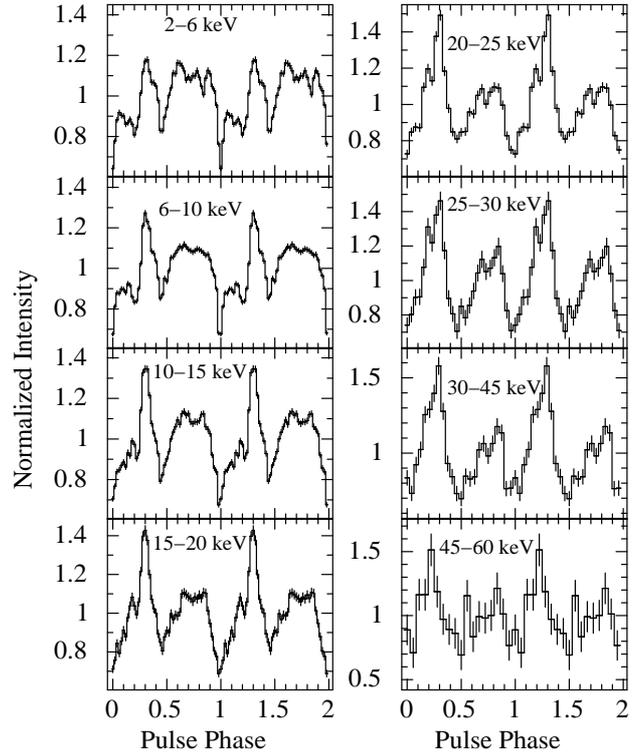}
\caption{Energy resolved pulse profiles of 2S~1417-624 obtained from 
the {\it RXTE} observation on 2009 November 9 (MJD 55144.28) at 
rising phase of the Type-II outburst. A double-peaked profile can be 
clearly seen in all the panels of the figure. The error-bars represent 
1$\sigma$ uncertainties. Two pulses in each panel are shown for clarity.} 
\label{fig3}
\end{figure}

\begin{figure}
\centering
\includegraphics[height=3.2in, width=3.9in, angle=-90]{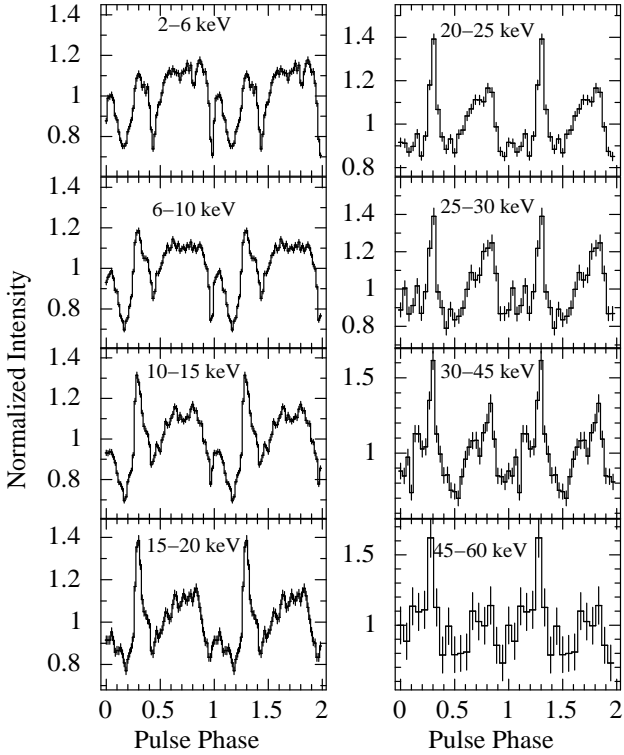}
\caption{Energy resolved pulse profiles of 2S~1417-624 obtained from 
the {\it RXTE} observation on 2009 November 17 (MJD 55152.20) near 
the peak of the Type-II outburst. A triple-peaked profile can be 
clearly visible in soft X-rays profiles that evolved into a double 
peaked structure at higher energies. The error-bars represent 
1$\sigma$ uncertainties. Two pulses in each panel are shown for clarity.} 
\label{fig4}
\end{figure}

We also investigated the evolution of pulse profile with energy at two 
different luminosities of the pulsar i.e. one during the onset of the 
outburst (MJD 55144.28) and the other at the peak of the outburst 
(MJD 55152.20). As described earlier, the source and background 
light curves in various energy ranges were extracted from the {\it RXTE} 
pointed observations on both the days, followed by barycenter correction 
on the background subtracted light curves and then folded with corresponding 
pulse period of the pulsar. Energy-resolved pulse profiles of the pulsar, 
obtained from both the observations are presented in Fig.~\ref{fig3} 
\& ~\ref{fig4}. We found that the pulse profiles of the pulsar are strongly 
energy dependent at different luminosities. The energy evolution of the 
first peak (0-0.5 phase range) in pulse profile was found to be faster than 
the second peak at late phases (0.5-1.0 phase range). Moreover, we also remarked 
that the energy evolution of pulse profile at lower luminosity was relatively simpler 
(Fig.~\ref{fig3}) as compared to the brightest observation at the peak of the 
outburst (Fig.~\ref{fig4}). In the latter observation, triple peaks are observed 
in the profile at energies below 15 keV (Fig.~\ref{fig4}). With increasing energy, 
the peak between 0.0-0.2 phase range gradually disappeared from the pulse profile 
in hard X-rays. We also noticed that the peak in 0.2-0.5 phase range was strongly 
energy dependent that evolved into a narrow component at higher energy, as seen in 
case of lower luminosity (Fig.~\ref{fig3}). The prominent evolution of this component 
led to a minor phase shift of $\sim$0.1 phase between the minima of soft (2-6 keV) 
and hard X-ray profiles (above 30 keV). As in case of the profiles at lower luminosity, 
a broad structure in 0.5-1.0 phase range was also detected. This component became narrower 
with energy though the peak intensity was almost constant across the energy ranges. 
Pulsations were detected up to $\sim$60 keV in both representative pulse profiles at 
different luminosities of the pulsar.

\begin{figure}
\centering
\includegraphics[height=2.9in, width=2.2in, angle=-90]{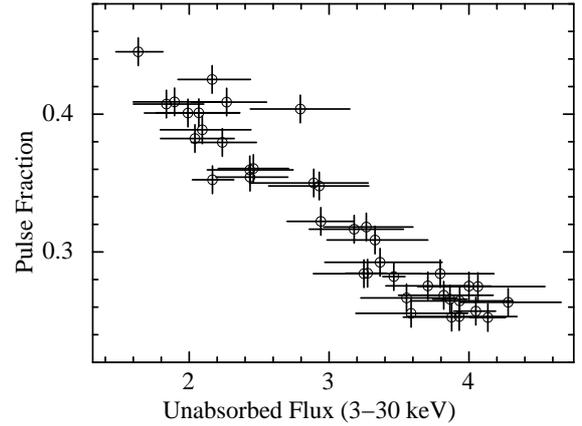}
\caption{Pulse fraction (ratio between the difference of maximum and minimum intensity 
and the sum of maximum and minimum intensity in the pulse profile of the pulsar) with 
the 3-30 keV unabsorbed flux (in 10$^{-9}$~erg s$^{-1}$ cm$^{-2}$ units) obtained from 
spectral fitting of the data from {\it RXTE}/PCA observations. The associated errors 
are calculated for 1$\sigma$ confidence level.} 
\label{fig5}
\end{figure}

In order to quantify the fraction of X-ray photons contributing 
to the pulsation, we estimated pulse fraction in 2-60 keV range   
for all the observations and shown in Fig.~\ref{fig5} along with 
the source flux. We calculated the pulse fraction by considering a 
ratio between the difference of maximum and minimum intensity and 
the sum of maximum and minimum intensity in the pulse profiles. A 
negative correlation between the pulse fraction of the pulsar and 
the unabsorbed source flux in 3-30 keV range can be clearly seen 
in Fig.~\ref{fig5}.

\subsection {Phase-averaged Spectroscopy}

\begin{figure}
\centering
\includegraphics[height=3.35in, width=2.7in, angle=-90]{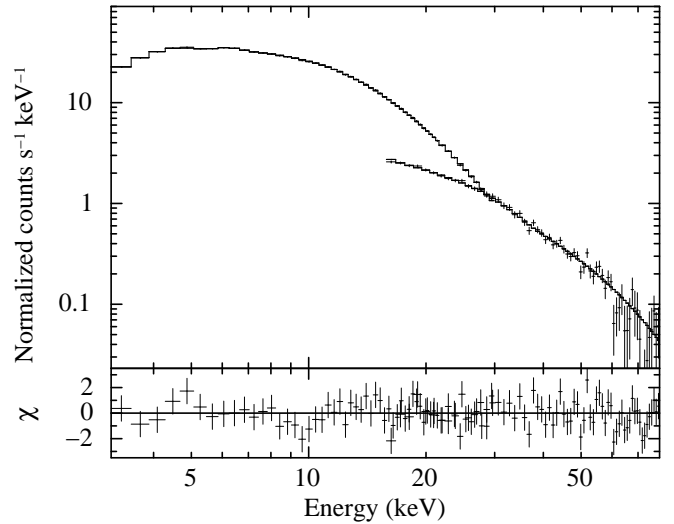}
\caption{Broad-band spectra of 2S~1417-624 in the 3-70~keV energy range (at the peak of 
the outburst; Obs-ID : 94032-02-03-03) and the best-fit model consisting of a cutoff 
power law model along with a Gaussian function at 6.4 keV for iron emission line are 
shown in the top panel. The bottom panel shows the contributions of the residuals to 
the $\chi^2$.} 
\label{fig6}
\end{figure}

We carried out pulse phase-averaged spectroscopy by using all available {\it RXTE} 
observations of the pulsar during the 2009 giant outburst. The source and background 
spectra were extracted by following the procedures as described in Section~2. Spectral
fitting was carried out by using \texttt{XSPEC} package of version 12.8.2. Data from PCA 
instrument (3-30 keV range) were used in our analysis. However, at the peak of the outburst, 
data from HEXTE instrument (15-70 keV range) were also used in our spectral fitting.  
As suggested by the instrument team, a systematic error of 1.0\% was added to the pulsar 
spectra obtained from the RXTE/PCA. Various conventional models such as negative 
and positive exponential cutoff (NPEX), Fermi-Dirac cutoff (FDCut), cutoff power law 
and high-energy cutoff power law continuum models were attempted to fit the 3-30 keV 
energy spectrum. Though a physical CompTT model was used to fit the spectrum, this model 
yielded a poor fit with a reduced-$\chi^2$~$\ge$2. A fluorescent iron emission line at 
$\sim$6.4 keV was also detected in the pulsar spectrum. It was found that an absorbed cutoff 
power law and high-energy cutoff power law models along with the Gaussian function for iron 
emission line fitted the spectra obtained from all the {\it RXTE} observations of the pulsar 
well. A representative broad-band spectrum in 3-70 keV range, obtained from the PCA and HEXTE 
data of the {\it RXTE} observation of the pulsar near the peak of the outburst (Obs. ID: 
94032-02-03-03) is shown in Fig.~\ref{fig6}. The spectral parameters obtained from fitting 
the 3-30 keV PCA spectra from all the available {\it RXTE} observations of the pulsar with 
the absorbed cutoff power law continuum model are presented in Table~\ref{table2}.

\begin{figure}
\centering
\includegraphics[height=3.35in, width=3.1in, angle=-90]{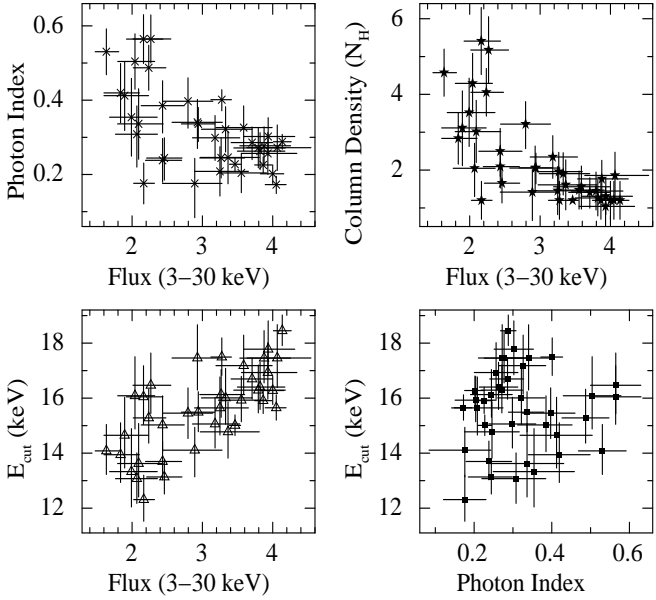}
\caption{Variation of spectral parameters such as power-law photon index, absorption
column density, cutoff energy with unabsorbed source flux in 3-30 keV range, obtained 
from the fitting of {\it RXTE}/PCA data with a cutoff power law continuum model. The 
error-bars are estimated for 90\% confidence level.} 
\label{fig7}
\end{figure}

Change in spectral parameters of the pulsar with the 3-30 keV unabsorbed flux, obtained 
from fitting the {\it RXTE} spectra with a cutoff power law continuum model are shown in 
Fig.~\ref{fig7}. The parameters such as power-law photon index and cutoff energy are 
found to show interesting variations with the 3-30 keV unabsorbed flux. The pulsar 
spectrum became hard with the increase in luminosity. At the same time, the cutoff 
energy was also increased with the increase in luminosity. Apart from the photon 
index and cutoff energy, the column density (N$_H$) which is variable in the range 
of (1-6) $\times$ 10$^{22}$ cm$^{-2}$ was found to be anti-correlated with the source 
flux. The estimated values of N$_H$ was found to be marginally higher than the value 
of absorption column density in the direction of the source. The cutoff energy did 
not show any dependence on the power-law photon index.

\begin{table*}
\centering 
\caption{Best-fitting spectral parameters of 2S~1417-624 obtained from fitting the 
{\it RXTE}/PCA data during the 2009 giant outburst with a cutoff power-law model. The 
errors are quoted for 90\% confidence level.}
\scalebox{0.9}{
\begin{tabular}{ |l | ccccccc}
\hline 
\hline 
Obs-id	  &N$_H$ $^a$           &Photon           &E$_{cut}$      &Line energy   &Eq. width     &Reduced-$\chi^{2}$   &Unabsorbed flux$^b$ \\
          &                     & index           &(keV)           &(keV)        &(eV)          &(d.o.f)              &(3-30 keV) \\
\hline \\
94032-02-01-00	&3.2$\pm$0.9	&0.41$\pm$0.09	&14.6$\pm$1.4	&6.52$\pm$0.27	&55$\pm$28	&1.17(43) &1.89$\pm$0.34	\\
94032-02-01-02	&3.2$\pm$0.6	&0.39$\pm$0.06	&15.4$\pm$0.9	&6.34$\pm$0.21	&128$\pm$60	&1.10(42) &2.79$\pm$0.35	\\
94032-02-01-06	&2.1$\pm$0.5	&0.23$\pm$0.05	&13.7$\pm$0.7	&6.43$\pm$0.15	&71$\pm$22	&0.50(47) &2.43$\pm$0.27	\\
94032-02-02-00	&1.4$\pm$0.6	&0.17$\pm$0.06	&14.1$\pm$0.8	&6.26$\pm$0.16	&133$\pm$78	&1.01(45) &2.89$\pm$0.38	\\
94032-02-02-01	&2.1$\pm$0.6	&0.34$\pm$0.06	&17.4$\pm$1.2	&6.35$\pm$0.18	&80$\pm$50	&0.84(48) &2.92$\pm$0.35	\\
94032-02-02-02	&1.9$\pm$0.6	&0.24$\pm$0.05	&16.1$\pm$0.9	&6.28$\pm$0.19	&60$\pm$23	&0.89(50) &3.26$\pm$0.33	\\
94032-02-02-03	&1.4$\pm$0.7	&0.20$\pm$0.06	&15.6$\pm$1.2	&6.32$\pm$0.12	&90$\pm$23	&0.67(49) &3.24$\pm$0.41        \\
94032-02-03-00	&1.5$\pm$0.5	&0.20$\pm$0.05	&15.9$\pm$0.9	&6.26$\pm$0.17	&71$\pm$22	&1.04(50) &3.55$\pm$0.35	\\
94032-02-03-01	&1.7$\pm$0.5	&0.27$\pm$0.05	&17.4$\pm$1.0	&6.38$\pm$0.15	&75$\pm$22	&1.15(50) &3.87$\pm$0.37	\\
94032-02-03-03	&1.1$\pm$0.2	&0.20$\pm$0.02	&16.3$\pm$0.5	&6.18$\pm$0.21	&120$\pm$60	&0.72(50) &3.99$\pm$0.15	\\	
94032-02-03-04	&1.1$\pm$0.2	&0.17$\pm$0.02	&15.6$\pm$0.5	&6.18$\pm$0.19	&103$\pm$50	&0.65(49) &4.04$\pm$0.14	\\
94032-02-03-05	&1.8$\pm$0.7	&0.27$\pm$0.06	&17.4$\pm$1.2	&6.39$\pm$0.17	&61$\pm$24	&0.85(49) &4.06$\pm$0.47	\\
94032-02-04-01	&1.4$\pm$0.3	&0.28$\pm$0.01	&18.4$\pm$0.6	&6.30$\pm$0.16	&98$\pm$37	&0.67(49) &4.13$\pm$0.12	\\
94032-02-04-02	&1.0$\pm$0.5	&0.25$\pm$0.05	&16.9$\pm$1.1	&6.28$\pm$0.15	&75$\pm$23	&0.71(46) &3.93$\pm$0.41	\\
94032-02-05-00	&1.4$\pm$0.4	&0.28$\pm$0.04	&16.7$\pm$0.7	&6.31$\pm$0.13	&69$\pm$31	&0.86(49) &3.70$\pm$0.29 	\\
94032-02-05-01	&1.3$\pm$0.5	&0.30$\pm$0.05	&17.8$\pm$1.0	&6.39$\pm$0.16	&67$\pm$22	&0.81(50) &3.93$\pm$0.37	\\
94032-02-05-02	&1.1$\pm$0.2	&0.22$\pm$0.02	&15.9$\pm$0.5	&6.29$\pm$0.21	&71$\pm$32	&0.57(49) &3.86$\pm$0.12	\\
94032-02-05-03	&1.1$\pm$0.5	&0.27$\pm$0.05	&16.3$\pm$0.9	&6.22$\pm$0.21	&66$\pm$23	&0.63(50) &3.79$\pm$0.38	\\
94032-02-05-04	&1.2$\pm$0.5	&0.26$\pm$0.04	&16.4$\pm$0.9	&6.31$\pm$0.11	&81$\pm$18	&0.79(49) &3.82$\pm$0.35	\\
94032-02-06-00	&1.5$\pm$0.6	&0.32$\pm$0.05	&17.2$\pm$1.1	&6.37$\pm$0.16	&82$\pm$42	&0.83(48) &3.58$\pm$0.4	        \\
94032-02-06-01	&1.0$\pm$0.4	&0.40$\pm$0.02	&17.5$\pm$0.7	&6.39$\pm$0.16	&64$\pm$39	&0.55(47) &3.27$\pm$0.13	\\
94032-02-06-02	&1.3$\pm$0.2	&0.22$\pm$0.01	&15.0$\pm$0.4	&6.45$\pm$0.12	&69$\pm$13	&0.88(48) &3.46$\pm$0.08	\\
94032-02-06-03	&1.9$\pm$0.5	&0.32$\pm$0.06	&16.1$\pm$1.1	&6.50$\pm$0.25	&55$\pm$24	&0.92(50) &3.32$\pm$0.37	\\
94032-02-06-04	&1.6$\pm$0.7	&0.24$\pm$0.07	&14.8$\pm$1.0	&6.24$\pm$0.14	&76$\pm$25	&0.79(49) &3.36$\pm$0.44	\\
94444-01-01-00	&2.0$\pm$0.4	&0.33$\pm$0.04	&15.5$\pm$0.7	&6.53$\pm$0.14	&68$\pm$30	&1.05(45) &2.94$\pm$0.23	\\
94444-01-01-01	&1.6$\pm$0.5	&0.24$\pm$0.05	&13.1$\pm$0.7	&6.44$\pm$0.16	&99$\pm$40	&0.91(48) &2.45$\pm$0.25	\\	
94444-01-02-00	&2.3$\pm$0.6	&0.29$\pm$0.06	&15.0$\pm$1.0	&6.39$\pm$0.16	&59$\pm$21	&0.53(47) &3.17$\pm$0.35	\\
94444-01-03-00	&2.5$\pm$0.6	&0.38$\pm$0.06	&15.0$\pm$1.1	&6.57$\pm$0.23	&72$\pm$52	&1.11(48) &2.43$\pm$0.3 	\\
94444-01-03-03	&2.0$\pm$0.7	&0.30$\pm$0.07	&13.1$\pm$0.9	&6.47$\pm$0.27	&101$\pm$64	&0.99(43) &2.06$\pm$0.29	\\
94444-01-04-00	&2.8$\pm$0.7	&0.41$\pm$0.07	&13.9$\pm$1.2	&6.59$\pm$0.32	&60$\pm$28	&1.27(45) &1.83$\pm$0.26	\\
94032-02-01-01	&4.3$\pm$0.8	&0.5$\pm$0.07	&16.1$\pm$1.4		&--		&--	&1.12(47) &2.04$\pm$0.28	\\
94032-02-01-03	&2.0$\pm$0.4	&0.17$\pm$0.05	&12.3$\pm$0.9		&--		&--	&1.02(45) &2.16$\pm$0.15	\\
94032-02-01-04	&3.5$\pm$1.0	&0.35$\pm$0.1	&13.3$\pm$1.6		&--		&--	&0.73(51) &1.99$\pm$0.36	\\
94032-02-01-05	&3.0$\pm$0.8	&0.33$\pm$0.09	&13.6$\pm$1.5		&--		&--	&1.04(47) &2.09$\pm$0.34	\\
94444-01-03-01	&5.1$\pm$0.2	&0.56$\pm$0.06	&16.5$\pm$1.2		&--		&--	&0.87(50) &2.26$\pm$0.28	\\
94444-01-03-02	&5.4$\pm$0.9	&0.56$\pm$0.06	&16.1$\pm$1.1		&--		&--	&0.85(50) &2.16$\pm$0.27	\\
94444-01-03-05	&4.0$\pm$0.7	&0.48$\pm$0.06	&15.3$\pm$1.0		&--		&--	&1.27(47) &2.23$\pm$0.24	\\
94444-01-04-01	&4.5$\pm$0.7	&0.53$\pm$0.06	&14.1$\pm$0.9		&--		&--	&0.86(47) &1.63$\pm$0.17	\\
\hline
\hline 
\end{tabular}}
\flushleft
$^a$ : Equivalent hydrogen column density in the unit of 10$^{22}$ cm$^{-2}$;  
$^b$ : in unit of 10$^{-9}$ erg s$^{-1}$ cm$^{-2}$. \\
\label{table2}
\end{table*}

\subsection*{Phase-resolved Spectroscopy}

To understand the emission geometry, cause of pulse profile evolution from double-peaked to
triple-peaked profile and investigate the changes in spectral parameters over pulse phases, 
we performed phase-resolved spectroscopy of the pulsar by using {\it RXTE} observations at 
rising phase and peak of the giant outburst. For this, the source and background spectra 
were extracted from GoodXenon mode data of PCA by following the standard procedure described  
in {\it RXTE} \textit{cookbook} and in {\it RXTE Guest Observer 
Facility}\footnote{https://heasarc.gsfc.nasa.gov/docs/xte/recipes/pulse\_phase.html}.
We selected two representative observations with Obs. IDs 94032-02-01-06 (rising phase 
of the outburst) and 94032-02-03-04 (peak of the outburst) to extract phase-resolved 
spectra. Using appropriate background and response files, the 3-30 keV spectrum 
for each phase bin was fitted with a high energy cutoff power law as well as cutoff 
power law continuum models along with a Gaussian function for iron emission line, as 
in case of phase-averaged spectroscopy. While fitting, the width of iron emission line 
was kept fixed at 0.1~keV. A systematic error of 1\% was added to the phase-resolved 
spectra. We found that both the models yielded acceptable fits with comparable 
spectral parameters. Phase-resolved spectroscopy of {\it RXTE} observations of
the pulsar at the rising phase and at the peak of the 2009 giant outburst, however,
did not yield any significant and systematic variation of spectral parameters over 
the pulse phases. The unabsorbed flux in 3-30 keV range was found to follow similar
pattern with pulse phase as that of the 3-10 keV and 10-30 keV pulse profiles
of the pulsar. Other parameters such as power-law photon index, cutoff energy
and absorption column density did not show any systematic variation with pulse
phase for both the observations. Therefore, it is difficult to draw any meaningful
conclusion from phase-resolved spectroscopy of the {\it RXTE} observations of the
pulsar during the 2009 giant outburst.


\section {Discussion and Conclusions}

We have presented a comprehensive spectral and timing analysis of the 
pulsar 2S~1417-624 by using {\it RXTE} observations during a giant outburst 
in 2009. The most intriguing aspect of our study is that, we witnessed 
peculiar changes in the pulse profiles of the pulsar for the first time. 
A double peaked profile at lower luminosity was found to evolve into a 
triple peaked structure with the increase in luminosity. This kind of 
variation has not been seen in the pulsar during any other Type~I and 
Type~II outbursts (\citealt{Finger1996, Inam2004}). During most of the 
earlier observations, pulse profiles were double-peaked and showed a 
marginal energy dependence. A simple broad profile was also reported 
at lower luminosity during the 1999 outburst \citep{Raichur2010}. This 
establishes the fact that the pulsar 2S~1417-624 showed luminosity dependent 
pulse profiles from a single peak to multiple peaks across a wide range 
of flux starting from 10$^{-11}$ to 10$^{-9}$ erg s$^{-1}$ cm$^{-2}$ 
(\citealt{Inam2004, Raichur2010}; and the present work). 

It has been already noticed that Be/X-ray binary pulsars such as EXO~2030+375 
(\citealt{Naik2013, Naik2015, Epili2017}), GX~304-1 \citep{Jaisawal2016} show 
multiple peaks in the pulse profile during outbursts. These peaks are known 
to be strongly energy dependent and usually associated with manifold minima 
or dip like structures in the soft X-rays. Phase-resolved spectroscopy of 
observations during X-ray outbursts of these transient pulsars showed the 
presence of additional matter at dip phases. This indicated that the dips 
in the pulse profiles are originated due to the absorption/obscuration of 
X-ray photons by narrow streams of matter that are phase-locked with the 
pulsar. As a result, the multiple peaks/dips observed in the pulse profiles 
of transient X-ray pulsars e.g. EXO~2030+375 and GX 304-1 are associated with 
the effect of partial absorption of the radiation beam rather than originating 
from different hot spots on the neutron star surface. 

In the present study, we have observed the evolution of a broad peak (in 0-0.5 
phase range; Fig.~\ref{fig2}) in the pulse profile into two peaks at a flux level 
of $\sim$3$\times$10$^{-9}$ erg~s$^{-1}$~cm$^{-2}$. This effectively evolved into 
three peaks at brighter phases of the giant outburst. Nevertheless, the source 
intensity was found to be relatively high i.e. $\sim$2.7$\times$10$^{-9}$ 
erg~s$^{-1}$ cm$^{-2}$ in 3-20~keV range during the 2009 outburst compared to
the previous events when the peak flux of the pulsar was observed to be 
$\sim$1.7$\times$10$^{-9}$ erg~s$^{-1}$ cm$^{-2}$ in above energy range 
(\citealt{Inam2004}). Therefore, a double peaked profile was only observed 
in the earlier studies. To explore the origin of multiple peaks at brighter phases 
of the pulsar, we have performed phase-resolved spectroscopy at two different 
luminosities of the pulsar. Our results showed moderate variation in the column 
density across the pulse phases in contrast to other Be/X-ray binary sources. 
Though, the observed column density (1-5 $\times$ 10$^{22}$~cm$^{-2}$) was found 
to be marginally higher than the Galactic value ($\sim$1.4$\times$10$^{22}$~cm$^{-2}$) 
along the line of sight, it is not sufficient to produce a remarkable dip or affect 
the beam function up to $\sim$20 keV, as seen in Fig.~\ref{fig4}. Limited low energy 
coverage of the {\it RXTE}/PCA (3 keV) may be one of the reason that constrained the 
column density measurement. Future studies with sensitive soft X-ray instruments 
such as {\it NICER} \citep{Gendreau2012} can explore the surrounding of the pulsar 
in depth. 

We have noticed distinct luminosity dependence of the pulse profiles during the 
2009 outburst. By assuming a source distance of 11~kpc (\citealt{Grindlay1984, 
Inam2004}), the 3-30 keV unabsorbed luminosity can be calculated to be in the 
range of $\sim$(2.4-6) $\times$ 10$^{37}$ erg~s$^{-1}$ during the 2009 Type-II 
outburst. Moreover, for a distance of 5~kpc, pulsar luminosity falls to a range 
of 4.8$\times$10$^{36}$ to 1.2$\times$10$^{37}$ erg~s$^{-1}$. For above distances, 
the source luminosity was estimated to be in the order of 10$^{37}$ erg~s$^{-1}$ 
which is a typical value of critical luminosity of accretion powered X-ray pulsars 
(\citealt{Becker2012, Reig2013, Mushtukov2015}). Critical luminosity of an accretion 
powered X-ray pulsar is crucial to understand the transition between two accretion 
regimes i.e. sub-critical and super-critical phases (\citealt{Basko1976, Becker2012}). 
At lower luminosity (sub-critical phase), accretion flow is expected to halt by Coulomb 
interaction close to the neutron star surface. In this regime, the high energy emission 
is dominated by a pencil beam (propagating parallel to magnetic field) or pulsed component 
\citep{Basko1976}. This component may produce a single or double peaked pulse profile 
depending on the visibility of the emission regions at poles of the neutron star. With 
increasing mass accretion rate, the radiation pressure takes lead and up-scatters the 
accreting plasma particles in a presence of radiation dominated shock above the surface.
X-ray photons emitted in this situation mostly diffuse through the side wall of the 
accretion column in the form of a fan beam pattern (perpendicular to magnetic field lines). 
This kind of pattern is un-pulsed and may produce a complex pulse profile. \citet{Becker2012} 
has calculated the critical luminosity by considering various physical processes in the 
accretion column which can be expressed as

\begin{eqnarray}
	L_{\rm crit} &=& 1.49 \times 10^{37}{\rm erg\,s}^{-1} \left( \frac{\Lambda}{0.1} \right)^{-7/5} w^{-28/15} \nonumber \\
	&&\times \left( \frac{M}{1.4{\rm\,M}_{\odot}} \right)^{29/30} \left( \frac{R}{10{\rm\,km}} \right)^{1/10} 
\left( \frac{B}{10^{12}{\rm\,G}} \right)^{16/15}
	\label{eqn:lcrit}
\end{eqnarray}

where $R$, $M$ and $B$ are the radius, mass and magnetic field (in 10$^{12}$~G) of the 
neutron star, respectively. The constant $\Lambda$ characterizes the mode of accretion 
which is assumed as 0.1 (for disk accretion) in our case. While the parameter $\omega$ 
represents the shape of the photon spectrum and considered as 1 \citep{Becker2007}. For 
standard case, above equation reduces to

\begin{eqnarray}
	L_{\rm crit} &=& 1.49 \times 10^{37} \left( \frac{B}{10^{12}{\rm\,G}} \right)^{16/15} {\rm erg\,s}^{-1}
	\label{eqn:lcrit-simplify}
\end{eqnarray}

Therefore, the critical luminosity for the pulsar 2S~1417-624 can be calculated to 
be $\sim$1.33$\times$10$^{37}$ erg$~s^{-1}$ by assuming a magnetic field strength
of 9$\times$10$^{11}$~G as reported by \citet{Inam2004}. The estimated value
of critical luminosity matches well with the pulsar luminosity observed during 
the 2009 giant outburst for considered distances. It also indicates that the 
pulsar was emitting at the critical luminosity or above the critical luminosity 
near the peak of the outburst. For sub-critical (below critical luminosity) sources, 
emission geometry is expected to be simple i.e. described by a pencil beam pattern
from pulsating component. Thus, the presence of double peaks with individual peaks 
separated by 0.5 pulse phase in the profile suggests a simple beam function originated 
from both the poles of the neutron star during outburst. As the source approaches the 
critical luminosity, the contribution from un-pulsed photons escaping through side walls 
of the column increases. This increase may influence the total beam geometry and produce
multiple peaks in the pulse profile, as seen in our study. This interpretation is further 
supported by the observed anti-correlation between pulsed fraction and source flux, as shown 
in Fig.~\ref{fig5}. A higher pulse fraction is expected from a pulsar in case of directly 
beamed emission. While the increasing contribution of the fan beam enhances the un-pulsed 
component or decreases the pulsed component at higher luminosity from the pulsar. This 
results a negative correlation in the pulse fraction with luminosity, as observed in our 
study. Other sources in luminosity range of 10$^{35}$--10$^{37}$ erg~s$^{-1}$ e.g. SXP 1323 
\citep{Yang2018} have also shown a negative trend with source intensity. Furthermore, the 
reflection from neutron star surface can also contribute to un-pulsed radiations at higher 
luminosity \citep{Mushtukov2018}. Therefore, we suggest that the changes in pulse morphology 
from a double to a triple peaked profile is associated with the change in beam pattern near 
the critical luminosity. 

In the support of results obtained from timing studies, we explored the spectral
properties of the pulsar by using {\it RXTE} data during the giant outburst. 
The 3-30 keV spectrum of 2S~1417-624 was found to be well described by a cutoff 
power law model along with a Gaussian function at 6.4 keV for the iron emission 
line. We studied changes in the spectral shape with the pulsar luminosity during 
the 2009 giant outburst. The photon index was found to be anti-correlated with the 
unabsorbed flux (see Fig.~\ref{fig7}), indicating that the spectrum became harder 
during the bright phase of the outburst. Interestingly, the observed power-law 
photon index during the 2009 outburst (in 0.1--0.6 range) is found to be lower 
than the earlier reported values (in 0.8--2.5 range) by \citet{Inam2004}. The 
authors also found a clear negative correlation in photon index with flux in 
(0.01-1) $\times$ 10$^{-9}$ erg~s$^{-1}$ cm$^{-2}$ range as in the present case. 
Along with negative correlation between the photon index and source flux, the 
values of photon index are found to be clustered in (3-4) $\times$ 10$^{-9}$ 
erg~s$^{-1}$ cm$^{-2}$ flux range.

The pattern of distribution of power-law photon index with source flux (Fig.~\ref{fig7}) 
has been observed in pulsars accreting in sub-critical regime or close to the critical 
luminosity (\citealt{Reig2013, Postnov2015, Epili2017} and references therein). The height 
of the pulsar emission region is considered to be in the range of a few kilometers 
\citep{Becker2012}. With increase in mass accretion rate, the height of the emission 
region gets reduced. At the same time, increase in the optical depth of the accretion 
column leads to the origin of harder spectra (or negative correlation) in the sub-critical 
regime. Nevertheless, the cutoff energy also showed a positive trend with the source flux. 
This parameter is associated with the plasma temperature and indicates the temperature 
enhancement of emission region with increasing luminosity (\citealt{Soong1990, Unger1992}). 
The presence of clustered values of photon index in (3-4) $\times$ 10$^{-9}$ erg~s$^{-1}$ 
cm$^{-2}$ flux range can be associated with the transition from sub-critical to super-critical
regimes during the giant outburst. Therefore, the results from our spectral analysis also 
support the idea of accretion transition close to the critical luminosity of the pulsar 
2S~1417-624 during the giant outburst. Accretion powered X-ray pulsars have strong magnetic 
field in the order of $\sim$10$^{12}$~G. Detection of absorption features in 10-100 keV 
range pulsar spectrum, known as cyclotron resonance scattering features, provides direct 
estimation of the pulsar magnetic field \citep{Jaisawal2017}. However, in the present case, 
no such features were detected in the pulsar spectra even at the peak of the giant outburst. 
Future observations with instruments covering broad energy range may provide an opportunity 
to constrain the magnetic field of the pulsar.

In summary, we have studied spectral and timing properties of 2S~1417-624 during 
a Type~II outburst in 2009. The pulse profile was found to evolve from a 
double-peaked profile to triple-peaked profile with the increase in the source 
luminosity. Spectral studies of the {\it RXTE} observations during the giant
outburst showed that the pulsar was accreting close to the critical luminosity
level. This led to the changes in the beam pattern, effectively from a pencil 
beam to a mixture of pencil and fan beams. This is also supported by variation 
in the pulse fraction with luminosity as well as results obtained from 
phase-resolved spectroscopy. We did not find any specific pattern in the column 
density with pulse phase of the pulsar indicating that the change in pattern
of the pulse profile is not associated with the inhomogeneous distribution
of matter (in the form of narrow accretion streams) around the poles. We suggest
that the three-peaked profile at brighter phases of outburst is caused due to the 
change in beam pattern at the critical luminosity in contrast to other effects 
such as absorption by a narrow stream of matter or emission from multiple hot spots on
the surface.

\section*{Acknowledgments} 
We thank the referee for his/her suggestions on the paper. The research work 
at Physical Research Laboratory is funded by the Department of Space, Government 
of India. GKJ acknowledges support from the Marie Sk{\l}odowska-Curie Actions 
grant no. 713683 (H2020; COFUNDPostdocDTU). This research has made use of data 
obtained through HEASARC Online Service, provided by the NASA/GSFC, in support 
of NASA High Energy Astrophysics Programs.

\def\apj{ApJ} \def\mnras{MNRAS}
\def\aap{A\&A} \def\apjl{ApJ} \def\aj{aj} \def\physrep{PhR}
\def\pre{PhRvE} \def\apjs{ApJS} \def\pasa{PASA} \def\pasj{PASJ}
\def\nat{Nat} \def\ssr{SSRv} \def\aapr{AAPR} \def\araa{ARAA}


\end{document}